# Realization of predicted exotic materials: The burden of proof

Oleksandr I. Malyi[1,2], Gustavo M. Dalpian[1,3], Xingang Zhao[1], Zhi Wang[1], and Alex Zunger[1*]

[1]Renewable and Sustainable Energy Institute, University of Colorado, Boulder, Colorado 80309
[2]Centre for Materials Science and Nanotechnology, Department of Physics, University of Oslo, P.O. Box 1048 Blindern, NO-0316 Oslo, Norway
[3]Centro de Ciências Naturais e Humanas, Universidade Federal do ABC, 09210-580, Santo André, SP, Brazil

[*]Corresponding author
e-mail: Alex.Zunger@Colorado.edu

## Abstract

Trove of exotic *topoloid structures* has recently been predicted by searching for compounds whose calculated band structure crossing points fulfill specific symmetry requirements. Discovery of exciting physical phenomena by experimental studies of such predicted compounds is just around the corner. Yet, the examination of some of these assumed high-symmetry structures suggests that not always will assembly of atoms in a configuration that yields exotic topological properties be protected against energy-lowering *symmetry breaking* modes. Indeed, although bulk topological characteristics lead to protected surface/edge states, nothing protects bulk states from structural instability. The burden of proof for theoretical predictions of exciting physical phenomena should include some compelling hints that such phenomena can live in thermodynamically stable (or near stable) compounds. Herein, we illustrate how the use of the calculated total (electron + ion) energies of candidate structures can remove false-positive predicted topoloids from the list of likely realizable compounds, to the benefit of the much-cherished iterative process of theory-experiment materials discovery.

## Introduction

One of the most exciting recent developments in condensed matter physics and solid-state chemistry is the discovery of topological behavior of matter[1-8] leading to unique transport properties at lower dimensions. Laboratory examination and future technological applications of the fascinating predicted properties of such topological phases naturally require identifying synthesizable and stable, topology-bearing compounds ("topoloids"). Searching for specific materials (characterized by **A**tomic identities, **C**omposition and **S**tructure, or **ACS**) that have required target property P (here, topological properties, but generally also novel photovoltaic semiconductors, transparent conductors, thermoelectric compounds or battery materials) generally follows an "Inverse Design" paradigm[9], illustrated for topological properties in Scheme 1. Given a compilation of materials with their characteristic ACSs, one first establishes the requisite theoretical conditions ("design principles") that would enable topological property P to exist in such crystal structures. Second, one examines if the selected compound and its computed band structure have the stipulated topological property *in the assumed crystal structure*. Finding a positive answer establishes potential *candidates for topological properties*. This approach has already provided thousands of specific predictions of topoloids[10-24] recently reviewed in Nature News[25]) that should be topological insulators (TI's), or topological crystalline insulators (TCI), Dirac and Weyl semimetals or unconventional quasiparticles. Experimental discovery of exciting new physical properties in these predicted compounds is a much-anticipated exciting prospect[25].

Yet, despite the recent predictions of thousands of topological compounds in the world, very few such compounds have been actually synthesized and proven to be topological. We note that the predicted topoloids were obtained by using just the upper part of the general flow in Scheme 1 denoted as "partial screening flow". The latter screening has been often based on the assumption that the crystal will take up the highest symmetry possible, not considering the possibility of energy-lowering symmetry breaking that obviates topological properties. Unfortunately, both experimental and theoretical databases of structures and properties of compounds sometimes omit information on magnetism, spontaneous defect formation, or how would the structure change when significant doping is required. Furthermore, in the case of *polymorphous networks*[26] (structures having a *distribution* of different local motifs), experimental X-ray structure determination often approximates the structure by an artificial, high symmetry primitive unit cell describing the macroscopic average configuration. Such "virtual structures" should not be used as input to electronic structure calculations[26]. All these factors can affect the assumed symmetry of the crystal structure and therefore the ensuing calculated wavefunctions, thereby misrepresenting the correct topological class.

Undoubtedly, more of the projected topoloids from the published lists of candidates[10-24] will be confirmed with time. However, we hold that the required next steps before offering theoretically predicted topoloid compounds for experimental synthesis and evaluation should involve the theoretical projection of their stability towards the formation of alternative competing phases, as well as symmetry breaking perturbations, following the "whole screening flow" of Scheme 1. Searching for materials with more conventional properties than topology, such as good photovoltaic absorber[27] or battery materials[28], generally, do not involve such an extreme dependence on the precise symmetry of different electronic states. Stability tests with respect to decomposition into competing phases are generally applied in these simpler search problems[29]. They usually result in a robust set of materials that have the target property and are stable in the structure that harbors these specific properties.

However, assembling atoms in an assumed configuration that yields the topology-promoting band structure energies does not necessarily protect against symmetry-breaking modes that lower the total energy, signaling instability of the assumed structure. Although bulk topological characteristics lead to protected surface/edge states, nothing protects bulk states from thermodynamic instability. These considerations were often ignored, in part because topological science emerged from field theory and (quantum Hall) theories of electron gases. In these approaches, the structural stability does not play much of a role, as the electron or hole gases are artificially confined by physical (kinetic energy) barriers rather than placed in the vicinity of ions as in molecules and solids. Often, hypothetical structures are assumed in such studies just to enquire what is possible "in principle", and what are the emerging trends. However, when a predicted structure turns out to be a high-energy phase, it is difficult to argue that the prediction is merely done to understand what could be possible, in principle, because the high energy by itself delivers a physics message. For example, the TI-ness property requires band inversion, i.e., occupation of previously empty anti-bonding bands and depopulation of previously occupied bonding states[23]. But this process cannot always be done with impunity because it has the potential of diminishing cohesion, resulting in a high-energy structure that might not exist, even in principle. The safe way for theoretical predictions should, therefore, involve verification of the instability of the potential candidates with respect to competing structures by using feasible methods, e.g., phase diagram, phonon spectrum. One might argue that it is trivial that an intrinsically unstable material might be difficult, if not unwise to synthesize. What we describe here are a few nontrivial physical mechanisms that will render a seemingly stable compound to an unstable one, outlining how could such a change be detected theoretically.

Fortunately, this type of assessment can be readily accomplished by using the sister quantity that goes hand-in-hand with the band structure calculation of such compounds – the total electron + ion energy of the said competing compounds. This needed next step of inspecting the *total energies*, following the already remarkable search for single-particle (band) energies that promise topological properties[10-24] would complete the *burden of proof* (as far as theory is concerned) for recommending for experimentation not only exotic, but also potentially realizable materials that host with impunity exotic properties.

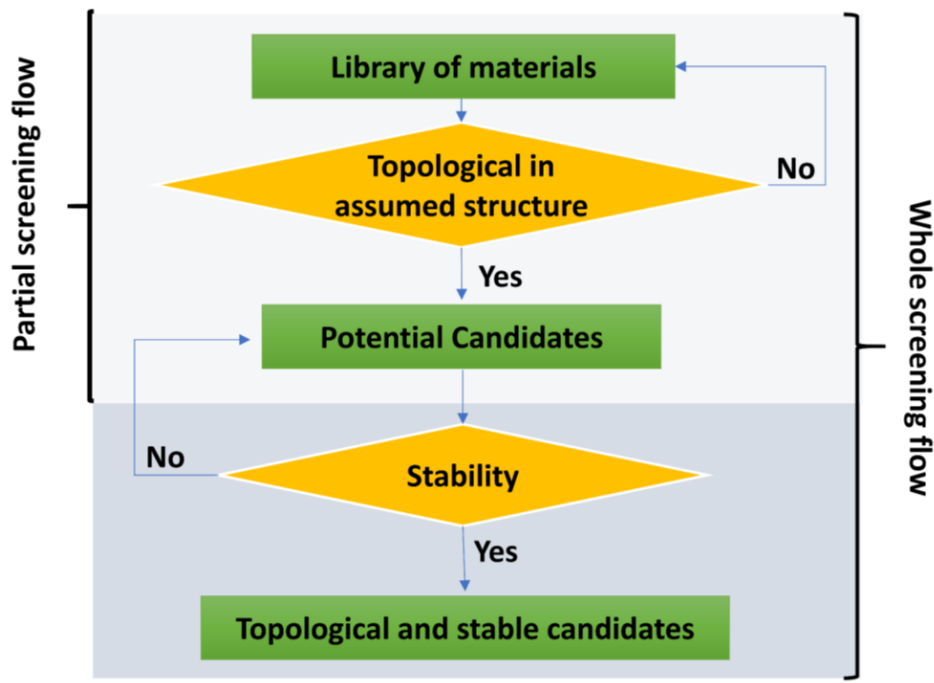

**Scheme 1:** Workflow for screening the potential and stable candidates with topological properties.

We illustrate a number of prototypical failure modes that can arise from restricting the predictions of topoloids to the examination of their band structures single-particle energies alone. We show how the use of the total energy[30] as an additional material selection filter might help avoid such false-positive predictions. The prototype modalities discussed are:

*(i) The hypothetical structure that host topology is unstable:* The selected crystal structure that manifests topology may not be the lowest energy structure. This is illustrated for the first predicted Dirac semimetal in three dimensions **BiO$_2$** in assumed cristobalite SiO$_2$ structure[19] which is shown here to be dynamically (phonon) unstable and disappears if its structure is allowed to relax to the energy-lowering stabler structure.

*(ii) Symmetry lowering by spontaneous defect formation defeats topology*: The newly predicted unconventional quasiparticle in **Ba$_4$Bi$_3$**[17] is destabilized by the spontaneous (energy lowering) formation of Bi vacancies that condense into ordered vacancy compounds whose low local symmetry defeats the specific degeneracy promoting topology.

*(iii) Symmetry lowering by magnetism (spin-polarization) removes the topology-promoting band degeneracy:* This is illustrated for the predicted 8-fold band degeneracy of *metallic* **CuBi$_2$O$_4$**[17] in assumed nonmagnetic configuration. Once magnetic spin-polarization is allowed, the total energy decreases, and the band degeneracy disappears.

*(iv) The predicted topological property requires doping that inherently destabilizes the topological structure:* This is illustrated for **BaBiO$_3$**[18] that requires an upshift of the Fermi energy by ~2 eV to meet the inverting energy bands. This occupation of massively antibonding states leads to a destabilizing increase in total energy relieved by structural transformations that defeat topology.

*(v) The topological property is caused by artificial band inversion due to systematic density functional theory (DFT) errors*: The method used to assess the band structure—DFT—requires an exchange-correlation energy functional. But current approximated functionals suffer from the self-interaction error that lowers the unoccupied bands and raises the occupied bands, causing, at the extreme, unphysical band inversion. This occurs in **InN** and **InAs**[10-13], well known normal insulators predicted by DFT with current functionals to be topological metals.

Avalible materials databases (e.g., ICSD[31]) contains hundreds of thousands entries. These are first pragmatically narrowed down by theorists[10-13] using filters such as avoid duplications, limit the maximum cell size to be examined, exclude theoretically difficult to calculate atoms (d or f electrons), exclude alloys and other structures with fractional occupation numbers, etc. After this narrowing down, the band structures of the remaining compounds are computed by density functional. For instance, Vergniory et al.[10] calculated 26,938 compounds from which 181 (34) were found to be topological insulators with DFT gap between occupied and unoccupied states of at least 0.001 (0.1) eV. The Swiss TOPOMAT database[13] uses different filters to narrow down the entries ending up with 13,628 directly inspected cases and found 50 (17) topological insulators with DFT band gaps of >0.001 (0.1) eV. We note, however, that the number of compounds and the numerical results in these online databases are regularly changed by the authors.

We have studied only a few compounds from the published lists of predicted topological insulators with finite band gaps. These already illustrate possible failure modalities that can be detected by using simple, additional "search filters" from the same density functional theory used to establish the requisite target band structure. The current list of predicted topological insulators with real band gaps might then be a diminishing domain.

**The approach**

Identifying actual realizations of topological compounds requires theories with a full atomic resolution that recognize site and space symmetries in crystals. The recent rebirth of density functional band theory of solids (and higher-order theories that use DFT to initialize the problem, such as DFT-DMFT[32], DFT-GW [33,34], DFT-QMC[35]) as a tool for uncovering topological behavior hidden in the spaghetti-like energy band structure, lies in DFT's ability to directly decode the consequences of an assumed ACS on the band structures. This affords a direct mapping of the theory onto the Periodic Table, via explicit incorporation of the electron-ion potential $V_{ext} = \sum -Z_\alpha/|r - R_\tau|$ in the single-particle problem, where $\{Z_\alpha\}$ are the atomic identities defined by atomic numbers or pseudopotentials, $\{R_\tau\}$ are the atomic positions with the appropriate periodic boundary conditions defined by the lattice vectors, and the sum extends over all atoms in the compound, thereby defining composition and stoichiometry.

This recent development of searching for ACS that host the conditions for topology[10-24] departs from the more traditional approaches to the search for compounds hosting other types of quantum behavior—unusual superconductivity, quantum spin liquids, or Kitaev unconventional magnetism. Such approaches[32,36,37] provide general guidance, specifying ranges of effective interactions in "Scenario Hamiltonians" where the new effects might be found. Such effective interactions, however, are not generally not mappable onto the Periodic Table, and thus do not commit to the chemical identity (ACS) of the compound hosting such interactions. This approach makes the process of identification of the system that can host such predicted "new physics" one of the successive guesses. Experimental verification—the heart and soul of natural sciences—is not easily accomplished.

To examine realizability of the predicted ACS that host topology, it may be wise to use not only the *band* single-particle energies as design filters but also the DFT total energy ($E_{total}$) and its derivatives with respect to displacements (forces, force constants) of the given crystal[30]. This total energy is given by

$$E_{total} = \sum_i \psi_i^*(r)(-\nabla^2)\psi_i(r)dr^3 + \sum_{i,\mu,l} \psi_i^*(r) U_{ps}^{(l)}(r-R_\mu)\hat{P}_l \psi_i(r)dr^3 + \frac{1}{2}E_{ee} + \frac{1}{2}E_{ion-ion} + E_{xc}. \quad (1)$$

Here, $\Psi_i^*(r)$ is (pseudo-)wavefunction of the valence electron where the index *i* denotes both the wavevector $k_i$ and the band index n and runs over all occupied valence states. $R_\mu$, $E_{ee}$, $E_{ion-ion}$, and $E_{xc}$ are the lattice vector, electron-electron Coulomb energy, lattice (ion-ion) energy and exchange-correlation energy, respectively. $U_{ps}^{(l)}(r-R_\mu)\hat{P}_l$ is the angular-momentum-dependent pseudopotentials, where $\hat{P}_l$ is the projection operator on angular momentum.

The need to use the total energy as an additional filter for material selection arises both because some candidate topoloid has been assumed to take up a hypothetical structure or composition not known to exist. Also, even if the selected structure does exist nominally, it is possible that simple microscopic mechanisms can create structural changes that defeat topology. Such situations are easily diagnosed by co-evaluation band structure conditions for topology, along with the potential instability with respect to total energy lowering deformations that defeat topology. The details of the methods used are described in the supplementary section S1.

**What about metastable compounds:** Certainly, metastable materials can be realized and are often used in functional applications today. Long-lived metastable structures protected by practically insurmountable activation barriers certainly exist in special structures. Compounds with the energy of roughly 50-100 meV/atom above the convex hull could be sometimes synthesized in a metastable form[38]: Metastability with respect to structural transformation is exemplified by diamond vs. graphite (diamond is 28 meV/atom less stable than graphite[39]); metastability with respect to a decomposition reaction is illustrated by $Ba_2YCu_3O_7$ which is 23 meV/atom above the convex hull and is predicted to decompose to $BaCuO_2$, $Ba(CuO)_2$, $Ba_2(CuO_2)_3$, and $Y_2O_3$[40]; metastability with respect to recombination of constituent elements is exemplified by nitrides where two nitrogen atoms can recombine to yield the highly stable $N_2$ molecule thus eliminating the parent compounds[38]. Indeed we and others generally consider theoretically compounds that are above the strict ground state ("Convex Hull") by reasonably small amounts of, say, less than 20-50 meV. The examples given above are rather gentle forms of metastability energy (smaller than kT during growth) and give rise to long-lived metastable structures protected by much higher activation barriers towards bond breaking. *The examples given in the current study, however, are not particularly gentle forms of metastability;* one would like to avoid such cases if synthesis would be attempted, as the pertinent target compounds are not likely to exist. We believe that seeking stable topoloid compounds from ordinary three-dimensional inorganic compounds is a good idea: metastable compounds in given structures may be difficult to make, in the first place, and if made, may create vulnerability towards decomposition during the inevitable perturbations (e.g., current, temperature) common in device applications

***Difference with respect to layer-by-layer growth:*** Certain specialized growth techniques grow a material under artificial conditions when it is stable. Then it is quenched to low temperatures when it is no longer stable but can be long-lived because activation energies for structural transformation can be large at such low temperatures. The classic example is layer-by-layer artificial growth from the gas phase by Molecular Beam Epitaxy (MBE), by opening and closing shutters that release certain atomic species and then quenched to low temperatures at which atomic diffusion is prohibitive. Predictions of the sequence of layers that delivers a given property, such as the spatial configuration of AlAs-GaAs having largest possible band gap[41], or the Si-Ge sequence that has the long-sought direct band gap[42], *can be done without considering global stability*, because the local stability is assured by the kinetically forbidden atomic diffusion. However, this condition is not applicable to conventional near-equilibrium melt growth or hydrothermal growth of *three-dimensional compounds* such as the topoloids predicted recently where layer-by-layer growth or quenching are not involved.

## *Results and Discussion*

### *A. False-Positive type 1: The hypothetical structure that host topology is thermodynamically unstable*

$BiO_2$ in the assumed cubic (Fd-3m) $SiO_2$ cristobalite crystal structure was predicted to be the first 3D Dirac semimetal (the special degeneracy point circled in Fig. 1a in red)[19]. However, Bi with its five $s^2p^3$ valence electrons may not have the type of bonding akin to fourfold valent Si ($s^2p^2$) even if Bi would disproportionate to $Bi^{3+}(s^2p^0)$ + $Bi^{5+}(s^0p^0)$. To examine this basic intuition, we calculated the Bi-O ground state (T=0) convex hull. This is done by evaluating the total energies of all conceivable known and hypothetical competing structures, examining the lowest energy structure at each composition and then identifying stable compositions that do not decompose into pairs of lower + higher compositions (Fig. 1b). Stable compounds lie on the convex hull whereas metastable compounds are located above it. Although a certain degree of metastability can be accepted as kinetically possible, the resulting ground state line shows that hypothetical $BiO_2$ structure is a high energy phase. The cristobalite structure is significantly higher in energy (263 meV/atom above the convex hull) than numerous non-cristobalite structures at lower energies (circles in Fig. 1b), so there is no apparent reason that it will form in that structure rather than any of the many lower energy phases. To evaluate the effect of finite temperature on the ground state diagram, we calculated the free energy at high temperature by using the machine learning model reported by Bartel *et al.*[43] (supplementary section S2), which includes the phonon entropy contribution. Our conclusion is that raising the temperature does not turn the cristobalite structure into a stable or near-stable structure.

Another important stability criterion is related to dynamic phonon factors. Some crystal structures or atomic configurations are reported in a configuration that have negative second derivatives (e.g., a maximum of the potential energy surface), so phonon calculations will lead to negative branches–a clear indication of dynamic instability. One can then use the method of Stokes and Hatch[44] to judge from the symmetry of the unstable phonons which alternative structures could replace the unstable one. Indeed, examination of the harmonic phonon dispersion curves (Fig. 1c) shows that this $SiO_2$ structure of $BiO_2$ is dynamically unstable, and therefore cannot exist as such. In the band structure of the phase that replaces the $SiO_2$-like polymorph (Fig. 1d), we observe that this material is, in fact, an insulator (see the gapped region in green) and so does not have the special degeneracy point akin to the predicted hypothetical cristobalite structure. We conclude that the cristobalite crystal structure of $BiO_2$ is not a likely stable compound, and the structures that are stable are not 3D Dirac semimetals.

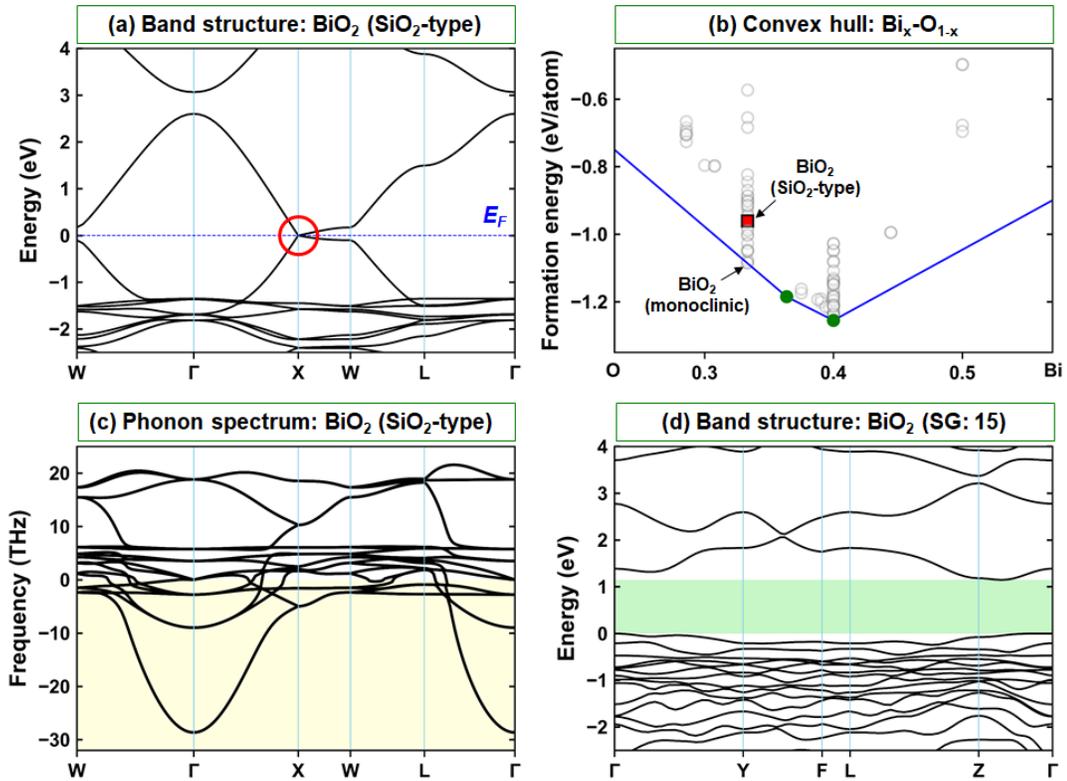

**Figure 1.** Electronic and phonon structure of the predicted first 3D Dirac semimetal $BiO_2$. (a) Band structure of the metallic cristobalite $SiO_2$-type, Fd-3m assumed structure showing the Dirac point (red circle). (b) Total energy convex hull of all phases showing that $SiO_2$-type $BiO_2$ is significantly (263 meV/atom) above the ground state line. (c) Phonon dispersion of the cristobalite structure, showing dynamically unstable modes (yellow highlights) and (d) band structure of the lowest energy monoclinic C2/c $BiO_2$ phase (space group 15) that is no longer a metal (gap in green highlight).

There are scores of other structures that were predicted to be topological in what was found to be chemically or thermodynamically unstable forms[20-22]. Among other channels, a structure can be unstable with respect to another structure type, or with respect to decomposition into its constituents. Both channels of instability as well as the effect of finite temperature (supplementary sections S1 and S2) can be examined[45]. One example is the prediction of new topological insulators with the highly desirable *light elements* ($Z < 50$) LiAgSe and NaAgSe in the assumed honeycomb ZrBeSi-type lattice[20]. On examination, it turns out that the ZrBeSi-type structure of LiAgSe and NaAgSe is not the ground-state structure[46]. According to total energy calculations performed on more than 50 potential crystal structures, LiAgSe should crystallize in a LiCaN structure while NaAgSe should crystallize in the $PbCl_2$-type structure. Furthermore, LiAgSe in the ZrBeSi-type structure is calculated to be unstable with respect to dissociation into $Li_2Se$ and $Ag_2Se$[46].

Lin *et al.*[22] predicted topological insulators among the ~2000 tested 18-electron half Heusler ABX compounds, assuming universally the cubic F-43m structure (α-type), finding exceptionally promising large inversion energies in 11 compounds, 8 based on I-Au-$X^{VI}$ (I= Li, Na, K, Rb and $X^{VI}$= O,S) and 3 involving Pd and Pt: LiPdCl, SrPtS, and BaPtS. However, the same DFT calculations extended to standard stability check[45] find that these latter compounds are unstable with respect to decomposition into their constituent binary compounds. Moreover, the assumed α-type structure[47] for the former group (Li, Na, K, Rb)Au(O,S) is not the lowest energy structure.

Another example is the predicted wide gap oxide TI $YBiO_3$[21]. Finding an oxide topological insulator that has a wide inversion gap has been sought for a long time, as it would overcome the inadvertent doping problem common in narrow gap TIs such as $Bi_2Se_3$ and could be integrated with the many oxide material functionalities such as ferroelectricity. Jin *et al.*[21] predicted through band structure calculations that $YBiO_3$ in the assumed $CaTiO_3$-type cubic perovskite structure (Pm-3m) with the Bi atoms located at the $O_h$ (1a Wyckoff position) would be a TI with an indirect gap of 0.18 eV and a direct gap of 0.33 eV. However, relaxing the lattice constant of $YBiO_3$[48] from the previously used 5.43 Å to the density-functional minimum energy equilibrium value for the same Pm-3m perovskite (a = 4.405 Å)[48] lowers the total energy by ~1 eV. Relaxing the condition that the structure is of the Pm-3m perovskite-type lowers the total energy by another ~1 eV. The resulting stable structure is not a TI.

### B. False-Positive type 2: Symmetry lowering by spontaneous defect formation defeats topology

$Ba_4Bi_3$ in space group 220 is an example of a material with special band structure degeneracy that corresponds to recently predicted unconventional quasiparticle[17] that remarkably has no analogy in particle physics. Its characteristic feature is the 8-fold degeneracy at the H point and 3-fold degeneracy at the P point of Brillouin zone (Fig. 2a, red circled ellipse). These degeneracies are a consequence of specific non-symmorphic symmetry existing in the space group 220. Fig. 2a constitutes the electronic structure description of the predicted unconventional quasiparticle.

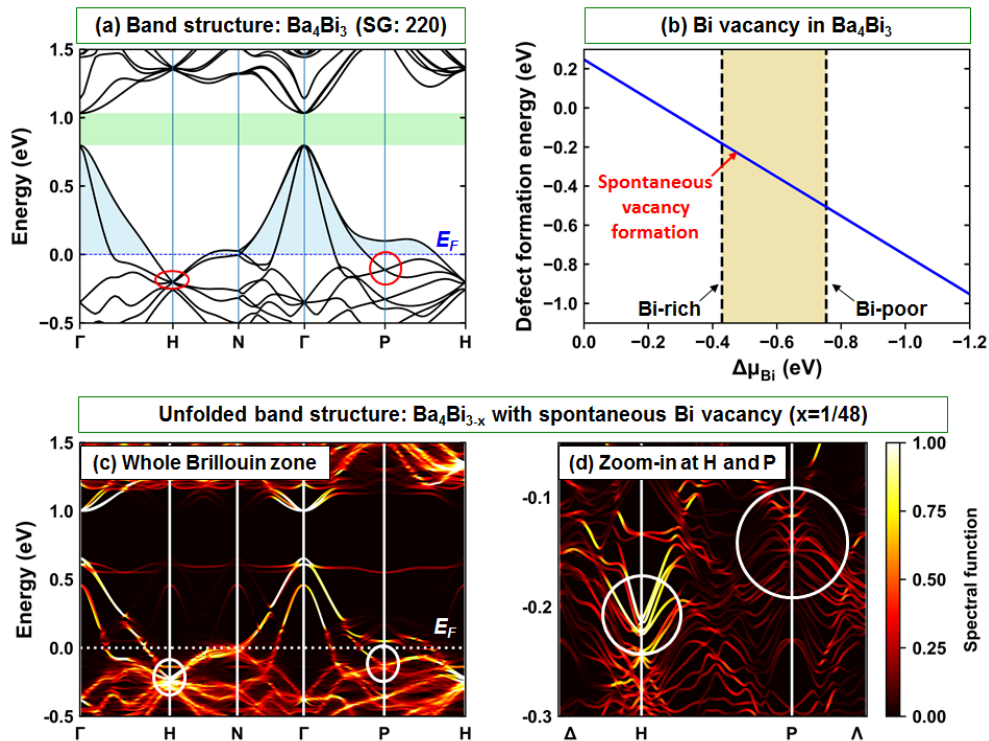

**Figure 2.** Spontaneous vacancy formation destroying band degeneracy in $Ba_4Bi_3$. (a) Band structure of ideal $Ba_4Bi_3$ in space group 220, showing the special degeneracy points (red circles) creating the topological properties of this predicted unusual quasiparticle. (b) Defect formation energy of Bi vacancy in $Ba_4Bi_3$ as a function of the Bi chemical potential. The chemical potential stability zone is shown in light yellow color, demonstrating the negative (i.e., exothermic) formation of Bi vacancies over a broad chemical potential range. The unfolded band structures for $Ba_4Bi_{3-x}$ containing about 2% of Bi vacancies is shown in (c) with a zoom-in shown in (d), indicating the disappearance of the special, topology-creating degeneracy. Supplementary section S3 shows the splitting of the Dirac cone as a function of the concentration of Bi vacancies.

The band structure of $Ba_4Bi_3$ (Fig. 2a) has a particularly unusual feature: a band gap (light-green shaded) above the Fermi level $E_F$, the latter being deep inside the valence band, indicating plenty of hole states (blue shaded region) in the valence band. This particulate electronic configuration might be unstable with respect to the spontaneous formation of intrinsic structural defects that compensate holes, i.e., defects that form electrons (donors). Such donors, formed above the Fermi energy would then transfer their electrons to the lower energy unoccupied hole states, thereby lowering the total energy of this defected structure relative to the ideal, perfect lattice. The defects formation energy—the total energy change with respect to the manipulatable chemical potentials of elements—can be used to estimate how hard is it to form the defects. For instance, the negative value indicates the exothermic reaction of forming vacancies spontaneously in thermodynamics, vice versa. The formation of low-temperature stable defects is known in other degenerate systems with internal band gap near the Fermi level, e.g., in ScS, $BaNbO_3$, $Ca_6Al_7O_{16}$[49,50]. Note that spontaneously formed defects are different from conventional defects that require energy for their formation and therefore exist in small quantities and only at elevated temperatures.

To examine this basic physical intuition, we calculated the total energy of the simplest donor defect namely Bi vacancy (see supplementary section S1 for details of defect calculations) in 16 formula units of $Ba_4Bi_3$ supercell (112 atoms). Fig. 2b shows the DFT calculated formation enthalpy of dilute Bi vacancy in $Ba_4Bi_3$ as a function of the Bi chemical potential, indicating that such defects would indeed form exothermically. The existence of dilute, stable defects further suggests the possibility of condensation of vacancies into ordered arrays (as in $Sc_2S_3$[50]). To examine this possibility, we have calculated the T=0 K stable phases of such structures by considering a replica of 16 $Ba_4Bi_3$ units with *p* Bi vacancies, searching via total energy minimization for stable and metastable configurations. The result detailed in supplementary section S3 shows that ideal $Ba_4Bi_3$ with holes in the valence band is a metastable structure with energy above the ground state convex hull, whereas $Ba_4Bi_{3-x}$ with vacancies is the ground state, as donor electrons relax into the low energy hole states.

The energy bands of the $Ba_4Bi_3$ structure with 2% of Bi vacancies are shown in Fig. 2c,d. To follow what happens to the special degeneracy points in the band structure of the ideal solid (ellipse in Fig. 2a) once vacancies are formed, we unfolded the 16-formula unit supercell band structure containing vacancies to the band structure in the primitive Brillouin zone. We see that upon the formation of even low concentration of Bi vacancies, the special degeneracies at the H and P points are removed, as the defects lower the symmetry of the lattice, hence also breaks the symmetry operators that protect the degenerate points ($\{C_{3,\bar{1}\bar{1}1}|0\frac{1}{2}\frac{1}{2}\}$, $\{C_{2y}|0\frac{1}{2}\frac{1}{2}\}$, $\{C_{2x}|\frac{3}{2}\frac{3}{2}0\}$, $\{IC_{4x}^-|\frac{1}{2}11\}$, $\{C_{2x}|\frac{1}{2}\frac{1}{2}0\}$, $\{C_{2y}|0\frac{1}{2}\frac{3}{2}\}$, $\{C_{3,111}^-|001\}$, $\{\sigma_{\bar{x}y}|\frac{1}{2}\frac{1}{2}\frac{1}{2}\}$). Supplementary section S3 shows the energy splitting of the two degenerate points as a function of vacancy concentration, demonstrating a sizeable splitting of over 100 meV is possible. We conclude that $Ba_4Bi_3$ in space group 220 structure is unlikely to retain this local symmetry needed to sustain the special 3-fold and 8-fold degenerate points because intrinsic, symmetry breaking vacancies would form exothermically. The experimental literature on $Ba_4Bi_3$ is very limited. Synthesis of $Ba_4Bi_3$ was achieved by direct reaction of the corresponding elemental by Li, Mudring and Corbett[51], who noted 1/9 random anion vacancies in this family of materials, and resistivity characteristic of a moderately poor metal, consistent with an anticipated one electron deficiency per formula unit for $Ba_4Bi_3$. These observations are entirely consistent with our picture of a non-degenerate electronic structure that may defeat topology.

### C. False-Positive type 3: Symmetry lowering by magnetism (spin-polarization) removes the topology-promoting band degeneracy: CuBi₂O₄

$CuBi_2O_4$ in the tetragonal P4/ncc crystal structure (space group 130) was predicted[52] to be a realization of an exotic *eightfold* metallic Fermion (Fig. 3a) owing to its specific Bravais lattice, that requires this degeneracy at the time-reversal invariant A point. The band structure (Fig. 3a) and the density of states (Fig. 3b) calculated[52] assuming a non-spin polarized configuration for the Cu atoms show a metallic behavior and an "intermediate band" where the degenerate Fermi energy resides. This constitutes the electronic description of the exotic eightfold metallic Fermion. However, the $Cu^{2+}$ ($d^9$) ion might be stabilized by developing a gap-opening spin-polarized magnetic structure.

To examine this basic physical intuition, we allow for magnetism in the form of different spin arrangements among neighboring Cu atoms, still in the space group 130. This is done by starting the calculations with a specific arrangement for the spins on the Cu atoms that will mimic ferromagnetism (parallel spins) or different antiferromagnetic arrangements (antiparallel spins). We find that an antiferromagnetic arrangement of the Cu spins can lower the total energy by as much as 498 meV/formula units. Different spin configuration (antiferromagnetic or ferromagnetic) lead to very similar total energies within a few meV/atom (see supplementary section S4). Breaking time-reversal symmetry through the inclusion of magnetic moments on the Cu atoms opens a large energy gap in the Fermi energy (green shaded area in Fig. 3c), splitting the metallic intermediate band into a valence band and conduction band with a DFT gap of 1.63 eV. Indeed, literature calculations and experiment[53] show that this material is an insulator with a large enough gap to be a potential photocatalyst. The paramagnetic phase is modeled using a supercell that allows moments to develop on individual Cu sites but keeps the total moment zero, as in the recent polymorphous approach to binary oxides[54]. Using a supercell with 756 atoms, with the same lattice parameters as the AFM phase (for simplicity), we find a band gap of 1.62 eV and a distribution of local magnetic moments around 0.63 Bohr Magnetons. This polymorphous representation of the paramagnetic phase is significantly lower in energy (by ~0.5 eV/formula unit) than the naïve approximation to a paramagnet as a nonmagnetic structure. We conclude that development of insulation-promoting magnetism in this system has such a significant energy lowering potential that it might very likely defeat the intended metallic state of the predicted new Fermion.

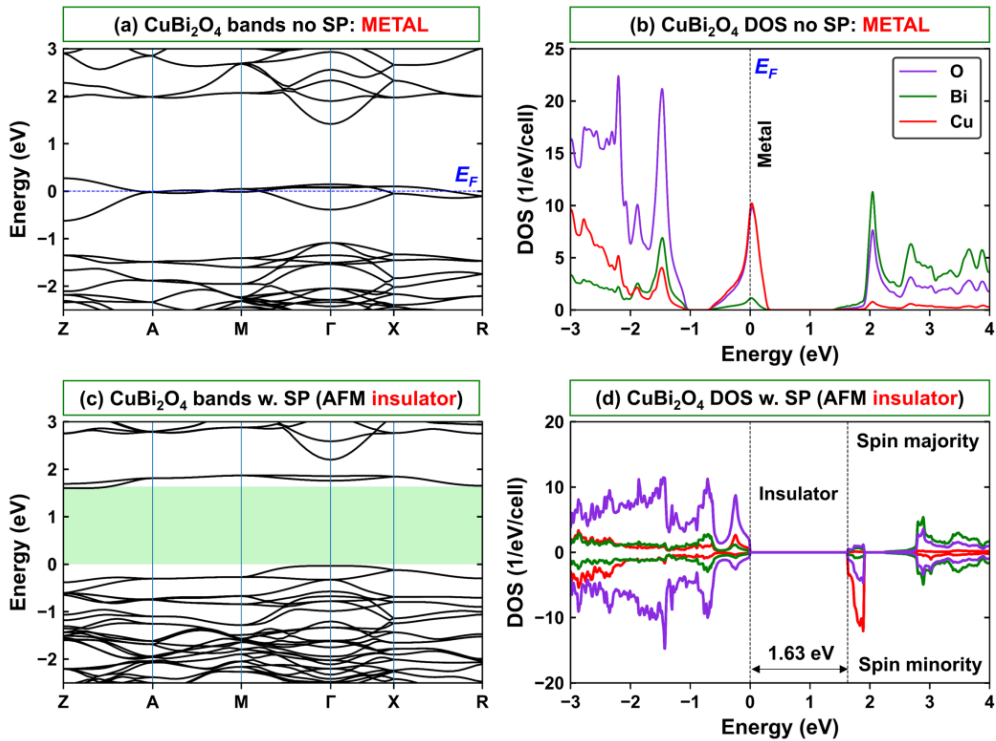

**Figure 3.** Spontaneous magnetism destroying topology-enabling band degeneracy in $CuBi_2O_4$. (a) Band structure and (b) atom-resolved density of states of $CuBi_2O_4$ in the tetragonal P4/ncc crystal structure (space group 130), constraining it to be non-spin-polarized, leading to a metallic solution. (c) and (d) are the band structure and density of states for the antiferromagnetic configuration that leads to the opening of the band gap and significant energy lowering. In (c), the spin-up and spin-down bands are degenerate.

**D. *False-Positive type 4: The predicted topological property requires doping that destabilizes the topological structure $BaBiO_3$***

Cubic (Fm-3m) $BaBiO_3$ was predicted[18] to be an exciting large gap oxide topological insulator, if it could be doped so as to shift up its natural Fermi level (zero of energy in Fig. 2a) by about 2 eV, where band inversion is predicted to occur (ellipse in Fig. 4a). Such doping is equivalent to one added electron per formula unit or doping density of $\approx 10^{22}$ electrons/cm$^3$. If one simulates doping by raising the Fermi level while disallowing the atoms to respond to such a perturbation, then indeed theory[18,55] would predict band inversion in the doped cubic solid (see supplementary section S5). However, doping is rarely a rigid-band event that does not provoke an electronic response[56]. Indeed, shifting $E_F$ to higher energy often leads at least to a self-regulating spontaneous formation of "killer defects", or at the extreme, to the decomposition of the host compound into stabler structures involving the dopant atoms[57]. In oxides, this is particularly true because of the limited ability of the polar lattice to screen such instabilities[58]. This can be examined by calculating the respective convex hull.

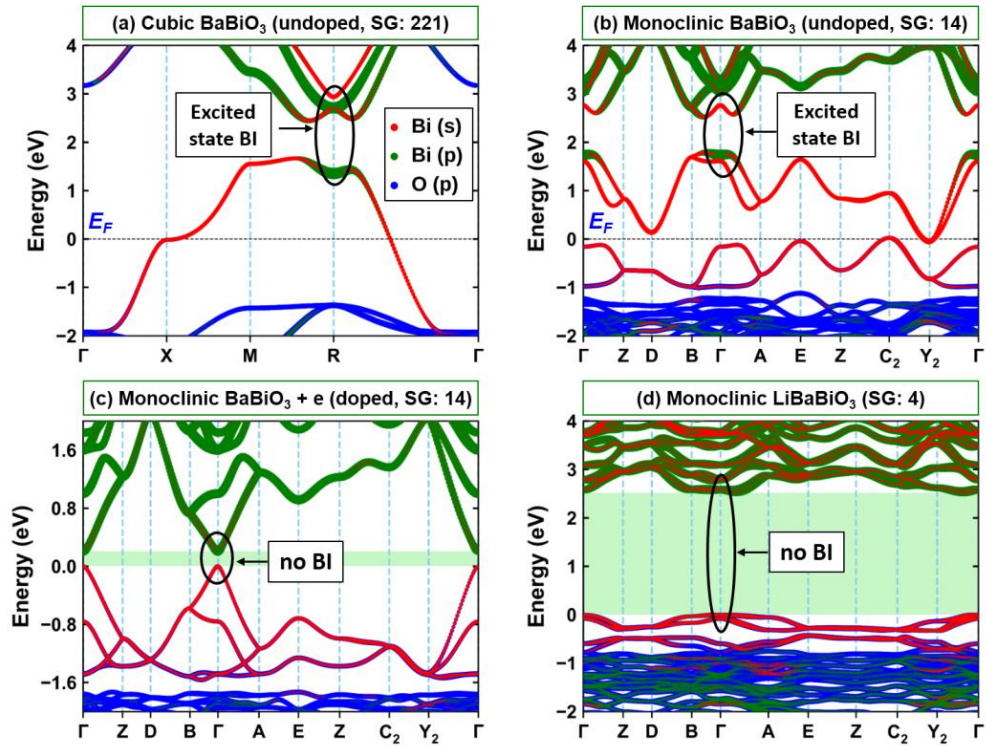

**Figure 4.** n-type doping destroying excited state band inversion in $BaBiO_3$. (a) Band structure of cubic $BaBiO_3$ showing band inversion (Bi p states in green below Bi s in red) at about 2 eV above the Fermi level. However, this cubic structure is unstable with respect to disproportionation of two of its $BiO_6$ octahedra into different (small vs large) octahedra, giving the known stable monoclinic structure (space group 14) that also has excited state (b) band inversion at about 2 eV above the Fermi level. This high energy band inversion is unstable with respect to n-type doping as shown for (c) electron-doped and (d) Li-doped monoclinic phases where a normal band order (Bi, s in red below Bi, p in green) is seen. Supplementary section S5 shows that the high energy band inversion is also not stable with respect to O-vacancy doping.

To examine this basic physical intuition, we calculated the equilibrium structure of the undoped as well as doped system by evaluating its total energy. Doping can result in the local asymmetric environment, lowering the thermal energy barrier for phase transition, leading to phase transition. We find first that the cubic (Fm-3m) phase of $BaBiO_3$ having a single $BiO_6$ octahedron is unstable with respect to disproportionation into two octahedra with different volumes often loosely described as $Ba_2[Bi^{3+} Bi^{5+}]O_6$, best described as the monoclinic structure (space group 14; $P2_1/c$), with metallic band structure shown in Fig. 4b. Once doped (by electron, by Li, or by O vacancies), this structure responds to the occupation of its previously empty conduction bands, rearranging its bonds, and leading to a fundamentally new band structure (Figs. 4c, d and supplementary section S5 showing also doping of the cubic phase) that is a *normal insulator, not a topological insulator*. A similar situation applies to cubic RbTlCl$_3$ and KBiO$_3$[58].

A more extreme example of a destructive response to doping is the recent suggestion[55] of replacing 1/3 of the O atoms by the nominal n-type dopant fluorine to shift $E_F$ to the point where a TI state is expected. This was attempted in a recent paper where following doping the band structure was monitored, but the total energy was not probed either with respect to nudging the atoms in BaBiO$_2$F or with respect to decomposition reactions. We find hypothetical cubic BaBiO$_2$F is highly unstable with respect to competing phases, specifically decomposes to ground state Bi$_2$O$_3$, Ba$_2$Bi$_2$O$_5$, and BaF$_2$ compounds with exothermic decomposition energy of 285 meV/atom. Recent experimental attempts to synthesize BaBiO$_2$F were unsuccessful, a failure attributed to the formation of different decomposition products[59]. We thus conclude that highly doped BaBiO$_3$ is unstable in the hypothetical structure that would have made it a topological insulator.

In general, one should be aware that shifting the Fermi energy at will via doping of quantum materials, as required in numerous theoretically predicted would- be- topoloids, is higly unlikely[56]. Such doping levels often lead to competing reactions that are thermodynamically favored over doping, and lead to products that generally lack the originally designed topological property.

### *E. False Positive type 5: The electronic Hamiltonian underestimates the band gap producing false band inversion in trivial insulators:*

Recent searches for topological phases[10-13] predicted that the Zinc blend (F-43m) form of InN and InAs are topological metals i.e., zero-gap topological insulators with inversion between the conduction s and valence p bands. This occurs because the initially empty metal-s $\Gamma_{6c}$ conduction band is lowered in energy for high atomic number cations such as In and Hg (due to the scalar relativistic Mass-Darwin effect). If this state is lowered sufficiently to get bellow the fourfold degenerate occupied P-type $\Gamma_{8v}$ valence band, then we have not only band inversion but also a topological metal since the Fermi energy must lie inside the half occupied $\Gamma_{8v}$ band. This is the actual case in HgTe[60]. Indeed, the PBE band structures with projection of the s and p states (Fig. 5a, c) for InN and InAs demonstrate that both systems have 0 eV band gap energies and electronic structures similar to that for HgTe topological material – being topological metals.

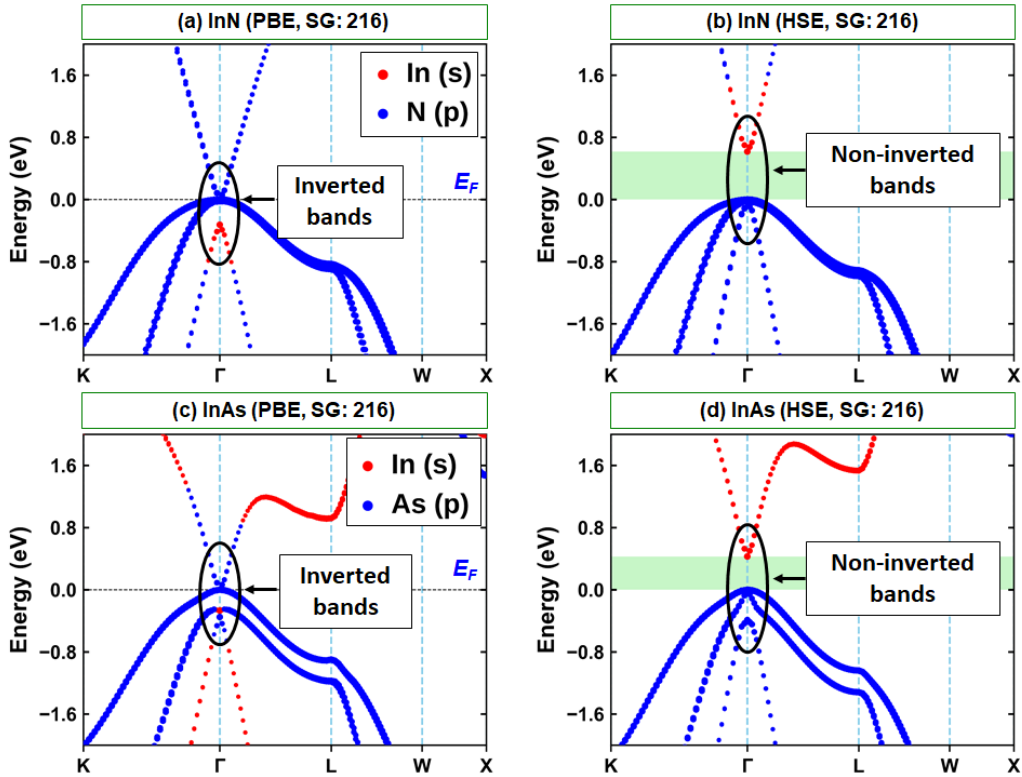

**Figure 5.** Self-interaction error resulting in artificial inverted band order. PBE band structures of (a) InN and (b) InAs showing band inversion in between In (s) (in red) and As/N (p) (in blue) states. This band inversion is caused by self-interaction error and disappears when a more accurate method to describe electronic structure is used. Band structures of (c) InN and (d) InAs computed using hybrid functional showing normal band order – valence band is defined by cation p states, while the conduction band is dominated by anion s states.

But DFT with current exchange-correlation functionals has the tendency to predict too low conductions bands and too high valence bands[61]. At the extreme, it could create an erroneous inversion of $\Gamma_{6c}$ and $\Gamma_{8v}$ if the initial band gaps are small. The question is therefore if the band inversion predicted in InAs and InN[10-13] is legitimate (as in HgTe), or false. To examine the effect of self-interaction error on the band order in the InN and InAs phases, we performed the hybrid functional calculations (see supplementary section 1) for both systems in comparison to the corresponding PBE calculations. Employing hybrid functional opens the band gaps for both compounds: the computed band gap energies for the InN and InAs phases are 0.62 and 0.43 eV, respectively. The orbital projected band structures (Fig. 5b, d) demonstrate that both InN and InAs phases are trivial insulators with normal band order – the valence band is dominated by cation p states, while anion s orbitals define the conduction band. Indeed, the existing experimental literature on the InN and InAs phases does not reveal any topological properties predicted within the simplified DFT calculations[62]. One should note that InN and InAs are not only the systems exhibiting such behavior. Specifically, the fictitious band inversion is found in InSb, GaSb, and CdO. However, it is not seen using the highly accurate electronic structure methods[63,64]. We conclude that search of realizable topological materials requires not only using the thermodynamic criteria to ensure in the materials stability but also ensuring that level of applied theory is sufficient to reproduce the physically correct band order for the proposed materials.

## Conclusions

Looking at instability modes of highly symmetric structures that harbor exotic properties reveals a number of basic symmetry breaking channels that can obviate exotic behavior. These additional filters are related to the stability of the hypothetical materials with respect to (i) decomposition to ground state crystalline compounds, (ii) spin polarization, (iii) formation of spontaneous intrinsic defects, (iv) external doping and (v) DFT underestimation of the band gap. The steps to account the above principles is straight forward: (i) perform analysis of thermodynamic stability of proposed materials via exploring energy convex hulls finding ground state compounds and stability of hypothetical material with respect to competing phases; (ii) verify stability of topological properties with respect to different spin arrangements and atomic/spin "nudging"; (iii) ensure that formation of point defects does not lead to spontaneous non-stoichiometry affecting the properties of the original materials; (iv) verifying that required doping is thermodynamically realizable and resulted system still possess the topology-enabling symmetry of the originally assumed structure. In addition to the above materials realization criteria, to propose a new topological system theoretically, (v) it is vital to ensure that topological properties of a hypothetical material are not affected by self-interaction error and more generally by the method used to describe electronic structure. Such additional material selection filters will help to avoid false-positive predictions and would complete the burden of proof for recommending to experimentalists not only exotic but also potentially realizable materials that host with structural impunity exotic properties. This would convert "theory of effects" to the theory of real structures that could harbor such effects with impunity. This would enhance the friendship (and credibility) between theorists with experimentalists.

**Acknowledgment**

Work at the University of Colorado Boulder was supported by the U.S. Department of Energy, Office of Science, Basic Energy Sciences, Materials Sciences and Engineering Division under Award No. DE-SC001046 and AFOSR, MURI program on material synthesis. AZ thanks Andrey Bernevig for a number of discussions on their results.

**Data availability**

All data needed to evaluate the conclusions in the paper are present in the paper and the Supplementary Materials. Additional data for this study are available from the corresponding author upon request.

**Appendix A.** Supplementary data are available